\documentclass[aps,pra,showpacs,amsmath,amssymb,amsfonts,tightenlines,twocolumn,lengthcheck]{revtex4-1}
\usepackage{amsmath}
\usepackage{amsfonts}
\usepackage{graphicx}
\usepackage{epsfig}
\usepackage{color}
\usepackage[colorlinks,citecolor=blue]{hyperref}

\begin{document}
\title{Optical hyperpolarization of heteronuclear spin singlet order in liquids}
\author{Y. Yang}
\address{ Key Laboratory of Low-Dimensional Quantum Structures and Quantum Control of Ministry of Education, Key Laboratory for Matter Microstructure and Function of Hunan Province, Department of Physics and Synergetic Innovation Center for Quantum Effects and Applications, Hunan Normal University, Changsha 410081, China}

\author{L. Zhou}
\address{ Key Laboratory of Low-Dimensional Quantum Structures and Quantum Control of Ministry of Education, Key Laboratory for Matter Microstructure and Function of Hunan Province, Department of Physics and Synergetic Innovation Center for Quantum Effects and Applications, Hunan Normal University, Changsha 410081, China}

\author{Q. Chen}
\email{qchen@hunnu.edu.cn} 
\address{ Key Laboratory of Low-Dimensional Quantum Structures and Quantum Control of Ministry of Education, Key Laboratory for Matter Microstructure and Function of Hunan Province, Department of Physics and Synergetic Innovation Center for Quantum Effects and Applications, Hunan Normal University, Changsha 410081, China}

\begin{abstract}
The nuclear spin singlet order involving coupled pairs of spins-1/2 may be used to store nuclear spin hyperpolarization in a room temperature liquid for a time much longer than the spin-lattice relaxation time $T_1$. There both are observations of long-lived homonuclear and heteronuclear spin-singlet order. Although hyperpolarized singlet order of the same species are accessible, hyperpolarized heteronuclear spin-singlet order has not been presented yet. Here we show hyperpolarized singlet order is achievable in the sample of $^{13}$C-labeled formic acid solution at room temperature by using optically polarized nitrogen vacancy (NV) center spins in nanodiamonds.
\end{abstract}
\date{\today}
\maketitle
\section{Introduction}
Singlet order has recently attracted considerable attention both experimentally ~\cite{Sheberstov2021,DeVience2012,Tanner2019} and theoretically ~\cite{Pileio2010(2),Levitt2019}, because it is long lived ~\cite{Carravetta2004}, silent and accessible on demand. Singlet order has been exploited for hyperpolarisation storage~\cite{Pileio2010,Pileio2017} and sensing applications such as sensing ligand binding~\cite{Buratto2016} or as a probe~\cite{Broadway2018} reporting on very slow flow~\cite{Pileio2015} and diffusion~\cite{Sarkar2008} over macroscopic length scales.

The existences of long-lived singlet order are demonstrated in homonuclear ~\cite{Tayler2012} and heteronuclear spins~\cite{Emondts2014}, as well as multiple-spin systems~\cite{Meie2013}. Hyperpolarized singlet order of the same species are accessible~\cite{Leggett2010,Carravetta2004a,Bornet2011,Pileio2013}, one intends to prepare hyperpolarized magnetization by using dynamic nuclear polarization (DNP)  and then convert it into singlet order. Several approaches are available such as preparing of singlet order in a high magnetic field by using a radiofrequency pulse sequence~\cite{Carravetta2004a} and applying an audio-frequency pulse sequence in a low magnetic field~\cite{Bornet2011}. In these methods, they may require extra hardware, time-consuming, low temperature and high magnetic field. Moreover, hyperpolarized singlet order is much more smaller than nuclear hyperpolarization, because singlet order is given by $p^2/3$ in which $p$ is the nuclear hyperpolarization~\cite{Tayler2012}. The practical method of DNP in solutions is Overhauser effect~\cite{Overhauser}, however it only works in low magnetic field limit and the reachable polarizations of different nuclear spins will be the same. Therefore the achievable singlet order is also given by $p^2/3$ which could be extremely low.

In this article, we intend to employ nitrogen-vacancy (NV) centers in nanodiamonds to achieve hyperpolarized heteronuclear singlet order in solutions at room temperature. The NV centers are excellent candidates as optically pumped hyperpolarisation agents~\cite{Abrams2014}, which allow for over 90\% electron spin polarization to be achieved in less than a microsecond by optical pumping while exhibiting a relaxation time in the millisecond range even at room temperature~\cite{Jelezko2004}. We consider a concrete setup consists of the nanodiamonds (NDs) immobilized in the hydrogel inside a flow channel~\cite{Q. Chen2016}. The spin singlet is formed by a strongly coupled $^{13}$C-$^{1}$H pair in $^{13}$C-labeled formic acid ($^{13}$CHOOH). Formic acid samples are used with mixtures of $^{13}$C-labeled formic acid and H$_2$O. The hydroxyl proton can be ignored because two- and three-bond couplings between the $^{13}$C-$^{1}$H system and the acidic proton can be neglected due to rapid exchange in protic solvents. Our method includes two procedures, the optical polarization built-up of the spin pairs by using optically pumped nitrogen-vacancy (NV) centers in nano-diamonds with the magnetic field $B=0.36$ T and singlet order generation of $^{13}$C-$^{1}$H pairs in formic acid sample when the magnetic field is adiabatic decreased to around $\sim10$ mG~\cite{Kiryutin2016,Rodin2020}. Singlet order hyperpolarization could be reachable at room temperature.
%%%%%%%%%%%%%%%%%%%%%%%%%%%%%
\begin{figure*}
\center\includegraphics[width=0.85\textwidth]{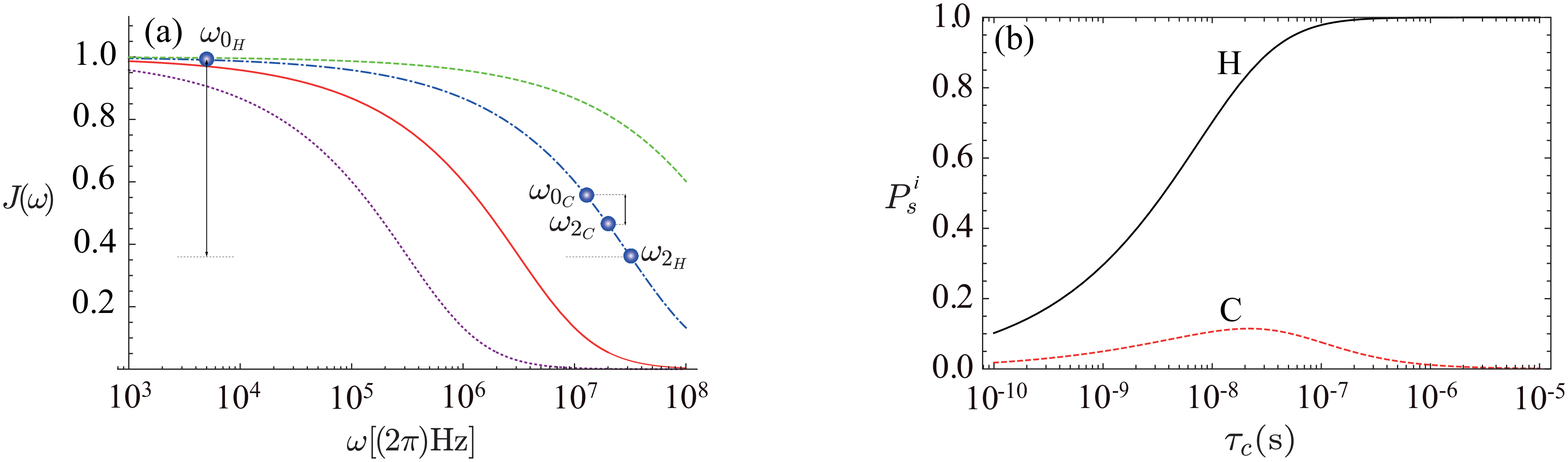}
\caption{(Color online) (a) Spectra density function $J\left( \omega \right)$ as a function of frequency $\omega $ and interaction correlation time $\tau _{c}$. The green dashed, blue dot-dashed, red solid, and purple dotted lines corresponds to $\tau _{c}=$1, 10, 100 and 1000 ns, respectively. The blue dots from left to right correspond to the spectral density functions of frequencies $\omega _{0_{H}}$, $\omega _{0_{C}}$, $\omega _{2_{C}}$ and $\omega _{2_{H}}$. (b) Nuclear spin polarization transfer rate $P_{s}^i$ changes with time $\tau _{c}$, which is defined by Eq.~(\ref{P_{s}}). The black solid and red dashed lines represent steady-state polarization of the $^{1}$H and $^{13}$C nuclear spins, respectively.}
\label{Fig1ab}
\end{figure*}
%%%%%%%%%%%%%%%%%%%%%%%%%%%%%
\section{Spin polarization built-up}
Formic acid samples in solutions are pumped into a channel containing a hydrogel polarizing cell with the immobilized nanodiamonds for nuclear spin polarization. The polarization target nuclear spins are coupled $^{13}$C-$^{1}$H pairs in $^{13}$C-labeled formic acid ($^{13}$CHOOH). The Hamiltonian of each pair is approximately described by
\begin{equation}
H_{CH}=g\vec{I}_{C}\cdot \vec{I}_{H}+(\gamma _{n_{C}}\vec{I}%
_{C}+\gamma _{n_{H}}\vec{I}_{H})\cdot \vec{B} ,
\label{A pair of Spin H}
\end{equation}
where the first term denotes as a mutual scalar coupling $g$ of $^{13}$C-$^{1}$H spin pairs and the second term represents the Zeeman interaction. Note that the scalar coupling $g=220$ Hz is much larger than the frequency of chemical shift.  $\gamma _{n_{C}}$ and $\gamma _{n_{H}}$ are gyromagnetic ratios of $^{13}$C and $^{1}$H nuclear spins, respectively. When a dc magnetic field is applied as $B=0.36$ T in the polarizing cell, the eigenstates are product states of the form $\left\vert \uparrow _{C}\uparrow _{H}\right\rangle$, $\left\vert \uparrow_{C}\downarrow _{H}\right\rangle$, $\left\vert \downarrow _{C}\uparrow_{H}\right\rangle$, $\left\vert \downarrow _{C}\downarrow _{H}\right\rangle$, the scalar coupling is negligible. 

The theory of spin polarisation via resonance-inclined transfer~\cite{Q. Chen2016} is employed to polarize $^{13}$C-$^{1}$H spin pairs in solution. Laser illumination is used for polarizing the NV spins and microwave (MW) irradiation facilitates the polarization transfer to nuclear spin pairs in samples. The steady-state polarization transfer of the nuclear spins is given by
\begin{equation}
P_{s}^i=-\frac{J\left( \omega _{0_i}\right) -J\left( \omega _{2_i}\right) }{%
J\left( \omega _{0_i}\right) +C_{0}J\left( \omega _{i}\right) +J\left( \omega
_{0_i}\right) },
\label{P_{s}}
\end{equation}
where $C_{0}=2\cot \varphi =2\frac{\epsilon }{\Omega }$ with $\Omega$ the Rabi frequency of the microwave driving and $\epsilon$ and the detuning of the MW field from the electron energy scale. The polarization built-up depends on the energy matching difference $\omega _{0_i}=\omega _{E}-\omega _{i}$ and mismatch $\omega _{2_i}=\omega _{E}+\omega _{i}$ between electron and nuclear frequencies $\omega _{E}$ and $\omega _{i}$. Note that the target nuclear spins could be $^{13}$C and $^{1}$H spins with the notation $i=C$  or $H$. $J\left( \omega \right)$ is the spectral density function 
$$
J\left( \omega \right) =\text{Re}\left[ \frac{1+\frac{\sqrt{i\xi}}{4}%
}{1+\sqrt{i\xi }+\frac{4\left( \sqrt{i\xi }\right) ^{2}}{9}+\frac{\left(
\sqrt{i\xi }\right) ^{3}}{9}}\right], 
$$
where $\xi =\omega \tau _{c}$, $\tau _{c}$ is a characteristic correlation time between the electron and nuclear spins, which could be adjusted via controlling the mesh sizes and types of hydrogels. The continual driving microwave field is given as $\Omega=\epsilon=8\sqrt{2}$ MHz. Thus as shown in Fig.~\ref{Fig1ab}(a), consider $\tau_c>10^{-8}$s, we have the spectral density function of  $^{1}H$ nuclear spin in the $^{13}$C-$^{1}$H pair in $^{13}$C-labeled formic acid $J\left( \omega_{0_H}\right)-J\left( \omega_{2_H}\right)\gg J\left( \omega_{0_C}\right)-J\left( \omega_{2_C}\right)$ corresponding to the $^{13}C$ nuclear spins. Therefore it is possible to achieve polarization transfer from NV centers to $^{1}$H nuclear spins, when polarization transfer to $^{13}C$ nuclear spins is suppressed. One has distinguished polarization differences in the $^{13}$C-$^{1}$H pair, see Fig.~\ref{Fig1ab}(b).

When $B=0.36$ T, as discussed in Ref. ~\cite{Chen}, the quantization axis of all NV centers is along the magnetic field, and the orientation of the symmetry axis of the NV center relative to the magnetic field is uniformly distributed over the unit sphere. The Hamiltonian of the NV center is given by~\cite{Chen}
\begin{eqnarray}
H_{NV} &=& (\gamma_{e}B+\delta(\theta))S_{z}+D(\theta)S_{z}^{2},
\end{eqnarray}
in which $\gamma _{e}$ is gyromagnetic ratio of eletron spin, $D(\theta)=\frac{D(1+3\cos(2\theta))+3E(1-\cos(2\theta))}{4}$, $\delta(\theta)=\frac{\gamma_e B|G_1|^2}{(\gamma_e B)^2-[D(\theta)]^2}+\frac{|G_2|^2}{2\gamma_e B}$, with $G_1=\frac{(D-E)\sin\theta\cos\theta}{\sqrt{2}}$, $G_2=\frac{D+3E+(E-D)\cos2\theta}{4}$, zero-field splitting $D=(2\pi)2.87$ GHz and the local strain $E=(2\pi)20$ MHz. When $B=0.36$ T ($\gamma _{e}B\gg D)$, the direction of the magnetic field defines the z-axis of the laboratory frame, $\theta$ denotes the angle between the NV center symmetry axis and the magnetic field axis. Clearly, the random orientations of the NV centers cause a variation of the zero-field splitting $D(\theta)$ across the entire interval $[-(2\pi)1.43\hspace{0.1cm}\text{GHz},\hspace{0.1cm}\text(2\pi)2.87\hspace{0.1cm}\text{GHz}]$ and $\delta(\theta)$ across the interval $[0 \hspace{0.1cm}\text{MHz},\hspace{0.1cm}(2\pi)140\hspace{0.1cm}\text{MHz}]$.  
%%%%%%%%%%%%%%%%%%%%%%%%%%%%%
\begin{figure*}
\center\includegraphics[width=0.85\textwidth]{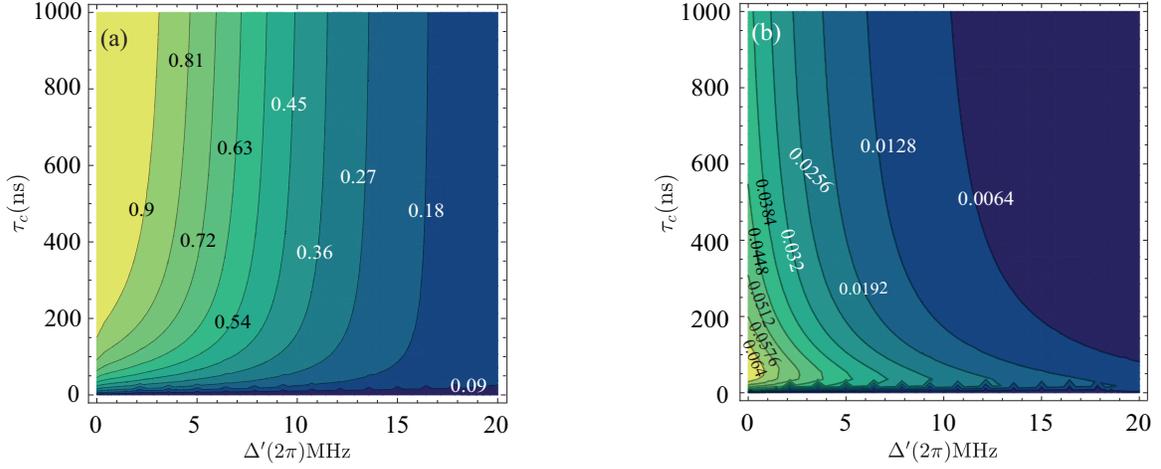}
\caption{(Color online)  (a) the steady-state polarization $P_{s}^{H}$ of $^{1}$H and (b) $P_{s}^{C}$ of $^{13}$C nuclear spins in the $^{13}$C-$^{1}$H pairs corresponds to the contour lines by considering $\epsilon _{0}=\Omega =\left( 2\pi \right) 8\sqrt{2}$ MHz and detuning from the resonance point $\Delta ^{\prime}=\epsilon \left(\theta \right) -\epsilon _{0}$. }
\label{Fig2}
\end{figure*}
%%%%%%%%%%%%%%%%%%%%%%%%%%%%%

When $\gamma _{e}B\gg D$, for an ensemble of randomly oriented nanodiamonds, the NV centers will be optically pumped to the state $|m_s=0\rangle_{\theta}$ that is defined by the relative orientation of the NV center with respect to the externally applied magnetic field~\cite{Chen}.  As discussed in Ref.~\cite{Chen}, two coordinate systems can be transformed into each other and employing of $ S_{z_{\theta}} = \cos\theta S_{z}-\sin\theta (\cos\phi S_{x}-\sin\phi S_{y})$, we can express the eigenstate $|m_s=0\rangle_{\theta}$ in the lab frame,
\begin{equation}
\label{optical}
|m_s=0\rangle_{\theta} = \cos\theta|0\rangle + \frac{\sin\theta}{\sqrt{2}}
(e^{i\phi}|+1\rangle-e^{-i\phi}|-1\rangle).
\end{equation}
If $\theta$ is large, the eigenstate $|0\rangle$ of the NV center in the laboratory frame differs significantly from the zero-field eigenstate $|0\rangle_{\theta}$ of the NV center. Hence optical initialization of randomly oriented NV centers lead to very different states depending on the orientation of the NV center. For $\theta\in[80^\circ,100^\circ]$,
the population in state $|0\rangle$ is very small and the initial state is well approximated by $\frac{1}{\sqrt{2}}(e^{i\phi}|+1\rangle-e^{-i\phi}|-1\rangle)$. In the subspace spanned by the states $\{|0\rangle, |-1\rangle\}$, the NV spin is well-polarized in state $|-1\rangle$ with polarization $P_{NV}\simeq 0.5$. So consider a continual optical pumping and the solid effect polarization mechanism, it's reasonable to have the initial polarization of the NV spin $P_e\simeq0.5^3=0.125$.

For utilizing the effect of resonance in the current setup, we tune the MW frequency to achieve resonant interaction for $\theta = 90^{\circ}$, and $\varepsilon_0=\varepsilon(90^\circ)=\Omega=(2\pi)8\sqrt{2}$ MHz for matching resonance of $\omega _{H}=(2\pi)16$ MHz. Due to the strong dependence of the energy splitting between the spin levels on $\theta$, for each NV spin with a specific $\epsilon(\theta)$, there is a detuning from resonance which leads to $\Delta'$ in Figs.~\ref{Fig2}(a) and~\ref{Fig2}(b). We can see that high steady-state polarization of $^{1}H$ nuclear spins is achieved for a wide range of detuning from the resonance, when the steady-state polarization of $^{13}C$ nuclear spins stay low, as shown in Figs.~\ref{Fig2}(a) and~\ref{Fig2}(b). Here we focus on the near resonant case around $\theta =90 ^\circ$ ($\Delta'<(2\pi)10$ MHz involves 5\% NV spins in Nanodiamonds). We estimate the efficiency of our scheme by calculating the average polarization rate of the solvent spins $\overline{ W}^i_{eff}= S^{-1}\int_{S}W_iP_e dS$, in which $P_e$ is the initial polarization of the corresponding NV spin according to the solid effect polarization mechanism~ and $S$ is the solid angle covered. The rate of the polarization transfer is given by $W_i= c_0(J(\omega_{0i})-J(\omega_{2i}))$, in which $c_0$ is a constant involving the nuclear properties of the interacting system~\cite{Q. Chen2016}.

\begin{figure}[h!]
\center\includegraphics[width=0.42\textwidth]{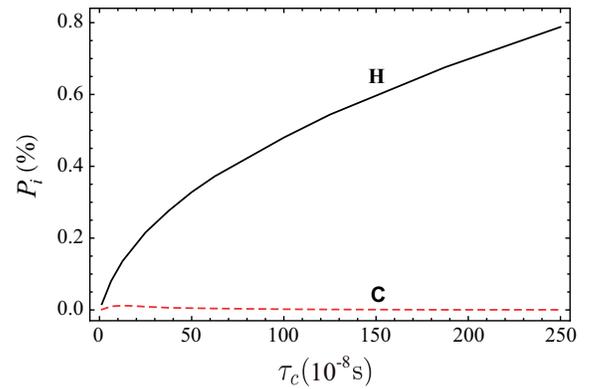}
\caption{(Color online) Considering the total polarization transfer time $t_1=1$ s, the difference in the total polarization can be obtained from the heteronuclear $^{13}$C-$^1$H pairs in the liquid sample calculated by Eq.~(\ref{p_{i}}). The black solid and red dashed lines present the possible total polarization of $^{1}$H nucleus and $^{13}$C nucleus, respectively.}
\label{Fig2Ps}
\end{figure}

To estimate the total polarization of the solvent spins, we use an approximate formula for the steady state bulk nuclear spin polarization of the solvent neglecting polarization diffusion,
\begin{eqnarray}
p_{i} &\approx&\frac{N_e}{N_i}\overline{W}^i_{eff}t_1.
\label{p_{i}}
\end{eqnarray}
Here $i=$ C or H denotes polarization built-up of $^{13}$C or $^1$H, $\frac{N_e}{N_i}$ is the ratio of the number of the NV spins to the $^{1}$H or the $^{13}$C spins in the polarization region, and $t_1$ is the time when the solvent is in contact with the hydrogel matrix. When the radius of the ND is 5 nm, suppose the flow channel is composed of a tube of a diameter of 1 mm, the size of the hydrogel layer is 1 mm and the resulting volume is filled with NDs of 10nm diameter such that they account for 12\% of the total volume. We choose the flow rate of $v = 10^{-3}$ m/s (this flow rate is achieved with commercial pumps, where flow rates over 10 times larger have been achieved in similar systems~\cite{ebert2012mobile}, and negligibly affects the molecular dynamics within $\tau_c$) such that $t_1 \sim 1 s$.  Formic acid samples are used with mixtures of $^{13}$C-labeled formic acid and H$_2$O (in the proportion 10:1). The hydroxyl proton can be ignored because two- and three-bond couplings between the $^{13}$C-$^{1}$H system and the acidic proton can be neglected due to rapid exchange in protic solvents. Given the density of spin pairs in the sample $13$ nm$^{-3}$, we obtain $N_e/N_H=N_e/N_C\sim1.6\times10^{-6}$, different polarizations of the $^{13}$C and $^{1}$H nuclear spins could be achievable in the solution. For example, when $\tau_c=15$ns, we have $p_{H}\approx 0.6\%\ll p_{C}$ as shown in Fig.~\ref{Fig2Ps}.
%%%%%%%%%%%%%%%%%%%%%%%%%%%%%
%%%%%%%%%%%%%%%%%%%%%%%%%%%%%

%%%%%%%%%%%%%%%%%%%%%%%%%%%%%
\begin{figure*}[tbp]
\center\includegraphics[width=0.55\textwidth]{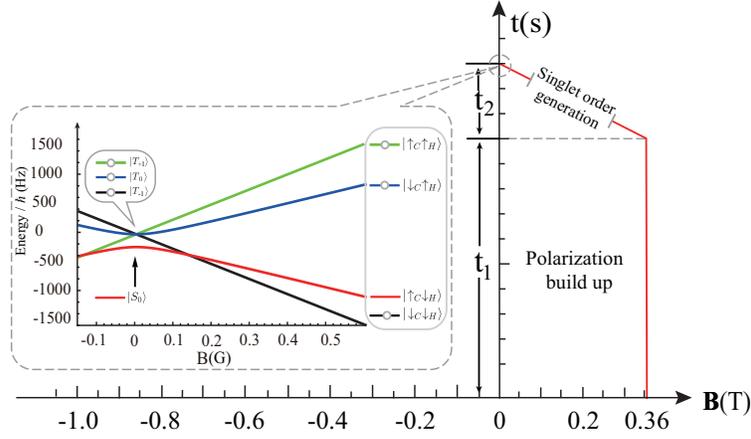}
\caption{(Color online)  The preparation of heteronuclear spin singlet order hyperpolarization includes two processes, spin polarization built-up and singlet order generation. The magnetic field strength changes from $0.36$ T to $10$ mG. Figure in dashed box considers energy levels for $^{13}$C-$^{1}$H heteronuclear spin pairs, in the presence of a scalar coupling with the coupling strength $g=220$ Hz. The eigenstates and eigenenergies of the system vary with the magnetic field (see Eq.~(\ref{A pair of Spin H})), and the population distribution is adiabatically transformed into the singlet and triplet eigenstates at zero magnetic field.}
\label{Fig3}
\end{figure*}
%%%%%%%%%%%%%%%%%%%%%%%%%%%%%
\section{Singlet order generation}
As shown in Fig.~\ref{Fig3}, the energy levels of the coupled spin pair change as a function of the magnetic field. According to Eq.~(\ref{A pair of Spin H}), when $\gamma _{n_{i}}B\gg g$, the eigenstates are product states of the form $\left\vert \uparrow _{C}\uparrow _{H}\right\rangle$, $\left\vert \uparrow_{C}\downarrow _{H}\right\rangle$, $\left\vert \downarrow _{C}\uparrow_{H}\right\rangle$, $\left\vert \downarrow _{C}\downarrow _{H}\right\rangle$. Around the level anti-crossing with $\gamma _{n_{i}}B\ll g$, the eigenstates are given by $\left\vert S_{0}\right\rangle$, $\left\vert T_{0}\right\rangle$, $\left\vert T_{+1}\right\rangle$, $\left\vert T_{-1}\right\rangle$. The correspondence between the low-field and high-field states is 
\begin{eqnarray}
\left\vert \downarrow _{C}\uparrow _{H}\right\rangle &\rightarrow&
\left\vert S_{0}\right\rangle =\frac{1}{\sqrt{2}}\left( \left\vert
\uparrow _{C}\downarrow _{H}\right\rangle -\left\vert \downarrow
_{C}\uparrow _{H}\right\rangle \right),  \nonumber \\
\left\vert \uparrow _{C}\downarrow _{H}\right\rangle &\rightarrow&
\left\vert T_{0}\right\rangle =\frac{1}{\sqrt{2}}\left( \left\vert
\uparrow _{C}\downarrow _{H}\right\rangle +\left\vert \downarrow
_{C}\uparrow _{H}\right\rangle \right),  \nonumber \\
\left\vert \uparrow _{C}\uparrow _{H}\right\rangle &\rightarrow&
\left\vert T_{+1}\right\rangle =\left\vert \uparrow _{C}\uparrow
_{H}\right\rangle,  \nonumber \\
\left\vert \downarrow _{C}\downarrow _{H}\right\rangle &\rightarrow&
\left\vert T_{-1}\right\rangle =\left\vert \downarrow _{C}\downarrow
_{H}\right\rangle.
\end{eqnarray}

Polarization of the zero-field levels could be achieved when the sample flows out of the polarizing cell and is collected in a region of low magnetic field (a solenoid generating a magnetic field of 10 mG), because the high-field eigenstates are adiabatically transformed into the nuclear singlet $\left\vert S_{0}\right\rangle$ and triplet $\left\vert T_{M}\right\rangle$ energy eigenstates in the low magnetic field, as shown in Fig.~\ref{Fig3}. Here the transit occurs as $t_2=0.3$ s to match the adiabatic condition to make sure that populations of the hyperpolarized high-field states are adiabatically transferred to those of the low-field states. When the sample flows out of the polarizing cell, the populations of the states of the nuclear pair are given by
\begin{eqnarray}
n_{\uparrow _{C}\uparrow _{H}} &=&\frac{1}{4}\left( 1+p_{C}\right) \left(
1+p_{H}\right) ,  \nonumber \\
n_{\uparrow _{C}\downarrow _{H}} &=&\frac{1}{4}\left( 1+p_{C}\right) \left(
1-p_{H}\right) ,  \nonumber \\
n_{\downarrow _{C}\uparrow _{H}} &=&\frac{1}{4}\left( 1-p_{C}\right) \left(
1+p_{H}\right) ,  \nonumber \\
n_{\downarrow _{C}\downarrow _{H}} &=&\frac{1}{4}\left( 1-p_{C}\right)
\left( 1-p_{H}\right).
\end{eqnarray}
Singlet order is defined by the difference between the population of the singlet state and the mean population of the three triplet states of the spin-1/2 pair in the low magnetic field. If relaxation losses during transport are neglected, singlet order is given by 
\begin{eqnarray}
p_{S}&=&n_{\uparrow _{C}\downarrow _{H}}-\frac{1}{3}\left( n_{\uparrow _{C}\uparrow _{H}}+n_{\uparrow _{C}\downarrow_{H}}+n_{\downarrow _{C}\downarrow _{H}}\right)  
\nonumber \\
&=&\frac{1}{3}\left( p_{C}-p_{H}-p_{C}p_{H}\right).
\end{eqnarray}
In our case by considering correlation time $\tau_c=15$ ns, there is a big polarization difference between the heteronuclear $^{13}$C-$^{1}$H spins in the spin pairs ($p_{H}\gg p_{C}$), $p_{S} \approx -p_{H}/3\approx-0.2\%$. Hyperpolarized singlet order is therefore available, when the sample flows out of the polarizing cell. The singlet polarization may be comparable to the longitudinal order and substantial compared to thermal polarization. The negative sign arises since strong polarization leads to an excess population in the triplet state, depleting the singlet state.
\section{Conclusion}
To summarize, we have shown that the hyperpolarized heteronuclear singlet order of $^{13}$C-$^{1}$H spin pairs in formic aid is available in the by using the NV centers in nanodiamonds at room temperature. The achieved singlet order is about a third of the longitudinal hyperpolarization of $^{1}$H spin, and we employ optical polarization of electron spins and all the procedures are implemented at room temperature without additional hardware and complexity involved in other singlet preparation methods.

\begin{acknowledgments}
Q. Chen is supported by Hunan Provincial Hundred People Plan (2019), Huxiang High-level Talent Gathering Project (2019) and Natural Science Foundation of Hunan Province, China (2019JJ10002); L. Zhou is supported by NSFC Grants No.11975095, No. 11935006 and the science and technology innovation Program of Hunan Province (Grant No. 2020RC4047).
\end{acknowledgments}

\end{document}